\documentclass[10pt,twocolumn,twoside,journal]{IEEEtran}
\IEEEoverridecommandlockouts
\usepackage{subfigure} 
\usepackage{graphicx}

\usepackage{amsmath,graphicx,amssymb,mathtools,bm}
\usepackage{subfigure}
\usepackage{hyperref}
\usepackage{cite}
\usepackage{amsmath,amssymb,amsfonts}  
\usepackage{textcomp}
\usepackage{xcolor}
\usepackage{verbatim}  
\usepackage{bm}  
\usepackage{mathrsfs} 
\usepackage{algorithmic} 
\usepackage{booktabs}
\usepackage{textcomp}  
\usepackage{multirow}  
\usepackage{lettrine}   
\usepackage{graphicx}  
\usepackage{color}  
\usepackage{amsmath}
\usepackage{amssymb}
\usepackage{stfloats} 
\usepackage{caption} 

\usepackage{color}  
\usepackage[ruled]{algorithm2e}

\begin{document}
	\newcommand{\tabincell}[2]{\begin{tabular}{@{}#1@{}}#2\end{tabular}}
	\newtheorem{Property}{\it Property} 
	
	\newtheorem{Proposition}{\bf Proposition}
	\newtheorem{remark}{Remark}
	\newenvironment{Proof}{{\indent \it Proof:}}{\hfill $\blacksquare$\par}

\title{Beamforming for Movable and Rotatable Antenna Enabled Multi-User Communications 
}
 
\author{Ruojing Zhao, Yifei Xu,  Songjie Yang,\\ Hua Chen,~\IEEEmembership{Senior Member,~IEEE}, Chadi Assi,~\IEEEmembership{Fellow,~IEEE}


\thanks{
Ruojing Zhao and Yifei Xu    contribute equally to this paper.
	
	Ruojing Zhao  and Yifei Xu are with the School of Telecommunications Engineering, Xidian University (e-mail: 932813155@qq.com; 1489855581@qq.com) 
	
	Songjie Yang is with the National Key Laboratory of Wireless Communications, University of Electronic Science and Technology of China, Chengdu 611731, China. (e-mail:	yangsongjie@std.uestc.edu.cn)
	
	 Hua Chen is with the Faculty of Electrical Engineering and Computer Science, Ningbo University, Ningbo 315211, China. (e-mail: dkchenhua0714@hotmail.com)
	 
	 Hua Chen is also with the Zhejiang Key Laboratory of Mobile Network Application Technology,  Ningbo 315211,  China.
	 
	 Chadi Assi is with Concordia University, Montreal, Quebec, H3G 1M8,
	 Canada (e-mail: assi@ciise.concordia.ca).}
	 
	}

\maketitle

\begin{abstract}
In the development of wireless communication technology, multiple-input multiple-output (MIMO) technology has emerged as a key enabler, significantly enhancing the capacity of communication systems. However, traditional MIMO systems, which rely on fixed-position antennas (FPAs) with spacing limitations, cannot fully exploit the channel variations in the continuous spatial domain, thus limiting the system's spatial multiplexing performance and diversity. To address these limitations, movable antennas (MAs) have been introduced, offering a breakthrough in signal processing and spatial multiplexing by overcoming the constraints of FPA-based systems. Furthermore, this paper extends the functionality of MAs by introducing movable rotatable antennas (MRAs), which enhance the system's ability to optimize performance in the spatial domain by adding rotational degrees of freedom. By incorporating a dynamic precoding framework based on both antenna position and rotation angle optimization, and employing the zero-forcing (ZF) precoding method, this paper proposes an efficient optimization approach aimed at improving signal quality, mitigating interference, and solving the non-linear, constrained optimization problem using the sequential quadratic programming (SQP) algorithm. This approach effectively enhances the communication system's performance.
\end{abstract}
\section{Introduction}   
Multiple-input multiple-output (MIMO) technology has emerged as a critical enabler, revolutionizing the communication field. By utilizing new spatial freedoms, MIMO significantly enhances the capacity of wireless communication systems, offering a novel approach to optimizing communication links and improving transmission efficiency \cite{1,THZ}. 

However, in both traditional MIMO systems and massive MIMO systems, transceivers typically rely on fixed-position antennas (FPAs), with antenna spacing not less than half a wavelength. This fixed, discrete antenna arrangement restricts the system's ability to fully leverage the channel variations in the continuous spatial domain, thereby limiting the diversity and spatial multiplexing performance of the MIMO system \cite{5}, \cite{6}. In practical applications, channel changes are continuous and complex, and traditional antenna deployment cannot fully exploit the rich channel variability in the spatial domain.

To overcome these limitations, fluid antenna systems (FASs) and movable antennas (MAs)  have been studied \cite{FA1,FA2,FA3,MA1,MA2,MA3,FAA2}. In \cite{MA1}, the authors maximized the multi-path channel gain under MA systems and showed the periodic behavior
of the multi-path channel gain in a given spatial field, providing insights for MA-enhanced communications.
The authors of \cite{FA1} derived a closed-form expression for the lower bound of capacity in FASs, confirming the substantial capacity gains derived from the diversity concealed within the compact space. Both \cite{MA1} and \cite{FA1} theoretically showcased the potential benefits of antenna position optimization in wireless communications. 
The study in \cite{FA4} explored point-to-point FAS communications using maximum ratio combining, revealing that the system's diversity order matches the total number of ports.
Moreover, \cite{FA2} noted in FASs that the multiplexing gain was directly proportional to the number of ports and inversely related to the signal-to-interference ratio target. Meanwhile, \cite{MA2,FA3} showed the potential of uplink power minimization through optimizing the user's antenna position. Conversely, \cite{MA3} explored MAs at the base station (BS) end, utilizing particle swarm optimization to optimize antenna positions with the objective of maximizing the minimum user rate.  In different perspectives, \cite{MA5} proposed flexible precoding based on sparse optimization for multi-user MA systems. Beyond these, MAs are promising for combining other wireless techniques, such as integrated sensing and communication \cite{MAISAC,MAISAC2}, mobile edge computing \cite{MAMEC}, physical layer security \cite{MASECURE}.

Antenna selection is another method to enhance the utilization of degrees of freedom in the continuous spatial domain. However, achieving high wireless transmission diversity through antenna selection requires a large number of antenna elements, resulting in significant costs  \cite{12}. In contrast, the MA system can achieve full spatial diversity with a smaller number of mobile antennas within a given area, making it a more cost-effective and practical solution.  Moreover, MAs could provide countinuous solutions for achieving excellent system performance.

Building upon the concept of MAs, this paper extends their functionality by incorporating the ability to rotate the antennas, resulting in movable rotatable antennas (MRAs). The rotation capability provides additional degrees of freedom in the spatial domain, enabling more comprehensive channel exploration and optimization of communication performance.  By incorporating both the position and rotation of the antennas as optimization variables, we aim to overcome the inherent limitations of FPAs and MAs, ultimately enhancing system performance. The introduction of antenna rotation allows for more precise control of signal alignment, further optimizing spatial multiplexing and diversity gains in the communication channel.

To this end, we propose a flexible precoding framework that dynamically adjusts both antenna positions and rotation angles to optimize the multi-user sum rate. The optimization problem is formulated with the antenna position and rotation as the key variables, which are jointly optimized to maximize signal quality and mitigate multi-user interference.
We employ the zero-forcing (ZF) precoding scheme to simplify the optimization process, as it decouples the interference between different users and makes the system's objective function dependent solely on the channel state information. This approach enables a clear and tractable optimization problem that focuses on antenna placement and rotation without introducing excessive complexity.
To solve this non-linear, constrained optimization problem, we utilize the sequential quadratic programming (SQP) algorithm, which is well-suited for handling such problems with both equality and inequality constraints.

{\emph {Notations}}:
  ${\left(  \cdot  \right)}^{ T}$, ${\left(  \cdot  \right)}^{ H}$, and $\left(\cdot\right)^{-1}$ denote   transpose, conjugate transpose, and inverse, respectively. $\vert\cdot\vert$ denotes the modulus.
$\Vert\cdot\Vert$  represents $\ell_2$ norm. 
$\Vert\mathbf{A}\Vert_F$ denotes the Frobenius norm of matrix $\mathbf{A}$.     $[\mathbf{A}]_{i,:}$ and $[\mathbf{A}]_{:,j}$ denote the $i$-the row and the $j$-the column of matrix $\mathbf{A}$, respectively.  $\nabla$ denotes the differential operator. $\odot$ denotes the Hadamard product. $\mathbb{E}\{\cdot\}$ denotes the expectation.   

\section{System Model}
  
As shown in Fig.1, we consider a MU-MISO downlink system, where the BS is equipped with $N\triangleq N_x\times N_z$ movable antennas distributed along the \(x\)-\(z\) plane, and serves \(K\) fixed single-antenna users. Multiple scatterers are considered to reflect signals between the BS and users.
  \begin{figure*}
	\centering 
	\includegraphics[width=6.5in]{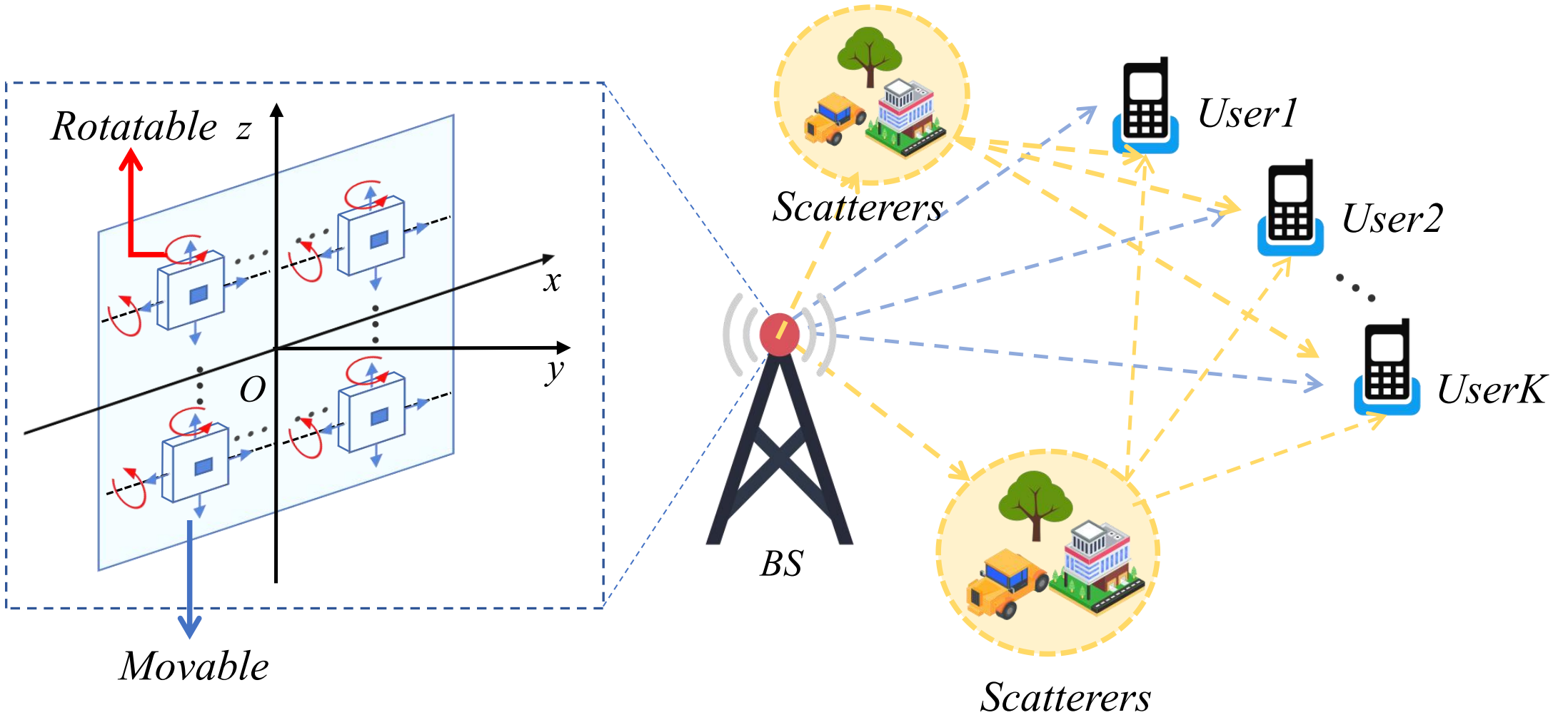}
	\caption{The multi-user communication scenario enabled by movable and rotatable antennas.}\label{SP} 
\end{figure*}  

Consider the downlink communication system, the received signal at the $k$-th user, $k\in\{1,\cdots,K\}$, can be expressed by
\begin{equation}
	y_k=\mathbf{h}_k^H\mathbf{F}\mathbf{s}+n_k,
\end{equation}
where $\mathbf{h}_k\in\mathbb{C}^{N\times 1}$ is the $k$-th user's channel, $\mathbf{F}\triangleq[\mathbf{f}_1,\cdots,\mathbf{f}_K]\in\mathbb{C}^{N\times K}$ is the precoding matrix,
$\mathbf{s}\in\mathbb{C}^{K\times 1}$ represents the $K$ data streams for $K$ users, $n_k$ represents the Gaussian additive white noise following $\mathcal{CN}(0,\sigma^2)$.

For the user $k$, the desired received signal is \( \mathbf{h}_k^H \mathbf{f}_k s_k \). Other users' signals will interfere with   user $k$ with the interference item \(\sum_{i \neq k}^K \mathbf{h}_k^H \mathbf{f}_i s_i \).
By assuming the i.i.d. transmit data such that $\mathbb{E}\{s_k^*s_k\}=1$ and $\mathbb{E}\{s_i^*s_k\}=0$, $\forall k , k\neq i$, then
the SINR of the $k$-user can be given by
\begin{equation}
	{\rm SINR}_k=\frac{\vert\mathbf{h}_k^H\mathbf{f}_k\vert^2}{\sum_{i,i\neq k}^{K}\vert\mathbf{h}_k^H\mathbf{f}_i\vert^2+\sigma^2}.
\end{equation} 

The spectral efficiency for user $k$ is given by
\begin{equation}\label{rk}
	R_k = \log_2 \left( 1 + \text{SINR}_k \right).
\end{equation}

In the following, we provide the spatial channel model, by assuming $L$ paths for all user channels, $\forall k$,
\begin{equation}\label{hk}
\mathbf{h}_k = \sqrt{\frac{1}{L}} \sum_{l=1}^L \beta_{k,l} \mathbf{a}(\theta_{k,l}, \phi_{k,l})\odot \mathbf{u}(\theta_{k,l},\phi_{k,l}),
\end{equation}
where $\beta_{k,l}$ is the complex path gain of the $l$-th path of the $k$-user's channel, $L$ is number of spatial channel paths,
$\phi_{k,l}\triangleq \cos(\varphi_{k,l})\sin(\vartheta_{k,l})$, and $\theta_{k,l}\triangleq\cos(\vartheta_{k,l})$ denote the virtual angles of the $l$-the path of the $k$-th user channel, and the array-angle manifold $\mathbf{a}(\theta_{k,l},\phi_{k,l})\in\mathbb{C}^{N\times 1}$ is given by \cite{MA5}
\begin{equation}
	\begin{aligned}
	&	\mathbf{a}(\theta_{k,l},\phi_{k,l})=\left[ e^{j\frac{2\pi}{\lambda}(\phi_{k,l} x_1+\theta_{k,l} z_1)},\cdots, \right. \\ & \ \ \ \ \ \ \ \ \ \ \  \left. e^{j\frac{2\pi}{\lambda}(\phi_{k,l} x_{n_x}+\theta_{k,l} z_{n_z})},\cdots, e^{j\frac{2\pi}{\lambda}(\phi_{k,l} x_{N_x}+\theta_{k,l} z_{N_z})} \right]^T,
	\end{aligned}
\end{equation}
where  $x_n$ and $z_n$ represent the position coordinates of the $n$-th antenna in the $x$-$z$ plane. This expression tells us the channel's phase will be impacted by the antenna position.

Additionally, the antenna pattern vector	$\mathbf{u}(\theta_{k,l}, \phi_{k,l})\in\mathbb{C}^{N\times 1}$ is given by
	\begin{equation}
		\begin{aligned}
			&\mathbf{u}(\theta_{k,l}, \phi_{k,l})\\ &= [
			u(\theta_{k,l}-\psi^\theta_1,\phi_{k,l}-\psi^\phi_1),\cdots, u(\theta_{k,l}-\psi^\theta_N,\phi_{k,l}-\psi^\phi_N)	]^T,
		\end{aligned}
	\end{equation}
	where $\{\psi_n^\theta,\psi_n^\phi\}_{n=1}^N$ denote the rotation angles for the $N$ antennas, and  $u(\theta,\phi)$ denotes antenna radiation pattern, which is expressed by the cosine pattern deployed on the $x$-$z$ plane \cite{FAA}.

	\section{Problem Formulation and Solution}
	
	Considering  the total power constraint, the spacing constraint which avoids the mutual coupling, and the maximal movable and rotation region,
	the sum rate maximization problem based on (\ref{rk}) is given by
	\begin{equation}\label{pro1}
	\begin{aligned}
&	\underset{\mathbf{F}, \bm{\Theta}}{\rm arg \ max} \  \sum_{k=1}^{K}R_k(\mathbf{F},\bm{\Theta}) \\
& 	{\rm s.t.} \Vert \mathbf{F} \Vert \leq P_t, \\ & \sqrt{(x_i-x_j)^2+(z_i-z_j)^2}\geq \frac{\lambda}{2}, i\neq j, \\ 
& \vert x_i \vert\leq x_{\rm max}, \vert z_i \vert\leq z_{\rm max},  \\
& \vert\psi_i^\theta\vert \leq \psi^\phi_{\rm max},  \vert\psi_i^\phi\vert \leq \psi^\theta_{\rm max},  \\
& \forall i,j\in\{1,\cdots, N\},
	\end{aligned}
	\end{equation}
where 	$\bm{\Theta}=[\mathbf{x},\mathbf{z},\bm{\psi}^\theta,\bm{\psi}^\phi]$ denotes the variables for antenna positions and rotation angles. $x_{\rm max}$ and ${z}_{\rm max}$ represent the maximal movable length along the $x$- and $z$-axis. $\psi^\theta_{\rm max}$ and $\psi^\phi_{\rm max}$ represent the maximal rotation angle around the $x$- and $z$-axis.

	Directly solving the problem in (\ref{pro1}) is difficult due to the highly coupled variables. Hence, we use the ZF precoding scheme to simplify the original problem by substituting the ZF precoder into (\ref{pro1}).  This also helps us directly evaluate the impact of antenna positions and angle rotations on the sum rate.
By stacking the channel vectors of all users in rows, we obtain $\mathbf{H}=[\mathbf{h}_1,\cdots,\mathbf{h}_K]^H\in\mathbb{C}^{K\times N}$. Then, the precoding matrix is given by
\begin{equation}
	\mathbf{F} = \mathbf{H}^H (\mathbf{H} \mathbf{H}^H)^{-1},
\end{equation}
followed by scaling \(\|\mathbf{f}_k\|^2 = \frac{P}{K}, \forall k\), with average power allocation. 
By substituting the precoding matrix into the SINR expression, the total sum rate expression simplified by ZF precoding can be obtained as follows:
\begin{equation}\label{rt}
R_{\text{total}} = \sum_{k=1}^K \log_2 \left(1 + \frac{P/K}{\sigma^2} \cdot \frac{1}{\left[((\mathbf{H}(\bm{\Theta})(\mathbf{H}(\bm{\Theta})^H)^{-1}\right]_{k,k}} \right),
\end{equation}
where $\mathbf{H}(\bm{\Theta})$ denotes the antenna channel function related to antenna positions and rotation angles in $\bm{\Theta}$.

Since this equation is exclusively dependent on the channel variable, it facilitates a straightforward analysis of antenna configurations.  This dependency enables an intuitive evaluation of how the spatial and angular flexibility of MRAs can be exploited to enhance channel conditions, thereby unlocking the potential for improved system performance.

Then, the problem in (\ref{pro1}) is reformulated by  
\begin{equation}\label{pro2}
	\begin{aligned} 
		\underset{\bm{\Theta}}{{\rm arg \ min}}
 & -\sum_{k=1}^K \log_2 \left(1 + \frac{P/K}{\sigma^2} \cdot \frac{1}{\left[(\mathbf{H}(\bm{\Theta})\mathbf{H}(\bm{\Theta})^H)^{-1}\right]_{k,k}} \right) \\
	&{\rm s.t.} \ \sqrt{(x_i-x_j)^2+(z_i-z_j)^2}\geq \frac{\lambda}{2}, i\neq j, \\ 
	& \vert x_i \vert\leq x_{\rm max}, \vert z_i \vert\leq z_{\rm max},  \\
	& \vert\psi_i^\theta\vert \leq \psi^\phi_{\rm max},  \vert\psi_i^\phi\vert \leq \psi^\theta_{\rm max},  \\
	& \forall i,j\in\{1,\cdots, N\},
	\end{aligned}
\end{equation}

This problem is also difficult to solve.
In order to solve the complex nonlinear optimization problem (\ref{pro2}) related to antenna positions, we use the SQP framework to transform the original problem into several simpler quadratic programming problems.
Firstly, the objective function of the nonlinear constraint problem formula (\ref{pro2}) is simplified to quadratic function at iteration point $\bm{\Theta}^t$ by Taylor expansion, and the constraint function is simplified to a linear function to obtain the following quadratic programming problem as follows:
\begin{equation}
	\begin{aligned}
&	\underset{\bm{\Theta}}{\rm arg \ min} \ R(\bm{\Theta}) = \frac{1}{2} (\bm{\Theta} - \bm{\Theta}^t)^T \nabla^2 f(\bm{\Theta}^t) (\bm{\Theta} - \bm{\Theta}^t)  \\  & \ \ \ \ \ \ \ \ \ \ \ \ \ \ \ \ \ \ \ \ + \nabla f(\bm{\Theta}^t)^T (\bm{\Theta} - \bm{\Theta}^t)\\
	&\ \ \ \ \ \ \ \ \ {\rm s.t.}\ \nabla g_{\ell }(\bm{\Theta}^{t})^T[\bm{\Theta} - \bm{\Theta}^{t}] + g_{\ell }(\bm{\Theta}^{t}) \leq 0, \\
	&\ \ \  \ \ \  \ \ \  \ \ \   \ \  \forall \ell \in \left\{1,\cdots, \binom{N}{2}\right\},
    \end{aligned}
\end{equation}
where $f(\bm{\Theta})$ is the objective expressed by (\ref{rt}), \( \nabla f(\bm{\Theta}^t) \) is the gradient of the objective function, and \( \nabla^2 f(\bm{\Theta}^t) \) is the Hessian matrix at the current point $\bm{\Theta}^t$. Additionally, the constraint \( g_{\ell }(\bm{\Theta}^{t})\in \{ \frac{\lambda}{2}-\sqrt{(x_i-x_j)^2+(z_i-z_j)^2} | i, j \in\{1,\cdots,N\} , i\neq j\} \) denotes the antenna spacing constraint.

In view of the fact that the current problem is only an approximation of the original problem, the solution obtained is not an accurate answer to the original problem, but only a better solution in the local area, but this solution can provide a guiding direction for the next iteration process, so in fact, we are seeking such a direction.

To solve this, we set \( \mathcal{S}^t=\bm{\Theta} - \bm{\Theta}^t \) and $V=\binom{N}{2}$ to organize the problem as follows for the variable $\mathcal{S}$:
\begin{equation}
	\begin{aligned}
\underset{\mathcal{S}}{\rm arg \ min}\ &\frac{1}{2}\mathcal{S}^T\nabla^{2} f(\bm{\Theta}^{t})\mathcal{S} + \nabla f(\bm{\Theta}^{t})^T\mathcal{S}\\
& {\rm s.t.} \ \nabla g_{\ell }(\bm{\Theta}^{t})^T\mathcal{S} + g_{\ell }(\bm{\Theta}^{t}) \leq 0,  \\
&\ \ \  \ \  \forall \ell \in \left\{1,\cdots, V\right\}.
   \end{aligned}
\end{equation}

However, this form still does not fully meet the standard formal requirements of quadratic programming, which is not conducive to us to directly use the related algorithms and theories of quadratic programming to solve it.

To this end, we define the following variables:
\begin{equation}
\begin{aligned}
	\mathbf{U} &= \nabla^{2} f(\bm{\Theta}^{t}), \\
	\mathbf{c} &= \nabla f(\bm{\Theta}^{t}), \\
	\mathbf{P} &= [\nabla g_1(\mathbf{\bm{\Theta}}^t), \nabla g_2(\mathbf{\bm{\Theta}}^t), \cdots, \nabla g_V(\mathbf{\bm{\Theta}}^t)]^T,\\
	\mathbf{b} &=-[g_1(\mathbf{\bm{\Theta}}^t), g_2(\mathbf{\bm{\Theta}}^t), \cdots, g_V(\mathbf{\bm{\Theta}}^t)]^T,
 \end{aligned}
\end{equation}
Then, the general form of the quadratic programming problem is given by
\begin{equation}\label{qp}
\begin{aligned}
\underset{\mathcal{S}}{\rm arg \ min}& \  \mathcal{S}^T \mathbf{U}\mathcal{S} + \mathbf{c}^T\mathcal{S}\\
&{\rm s.t.} \ \mathbf{P}\mathcal{S} \leq \mathbf{b}.\\
\end{aligned}
\end{equation}
 
To solve this quadratic programming problem, we take its optimal solution \( \mathcal{S}^\star \) as the next search direction \( \mathcal{S}^t \) of the original problem, and then expand the constrained one-dimensional search in this direction against the objective function of the original constraint problem, and obtain an approximate solution \(\bm{\Theta}^{t+1}\) of the original constraint problem. Repeating this process, the optimal solution of the original problem can be found.
Once the quadratic subproblem is solved and the optimal direction \( \mathcal{S}^t \) is found, we update the solution as follows:
\begin{equation}
\bm{\Theta}^{t+1} = \bm{\Theta}^t + \mathcal{S}^t.
\end{equation}
This update moves the current parameter \(\bm{\Theta}^t\) in the direction of \( \mathcal{S}^t \), which minimizes the objective function locally.

After updating \( \bm{\Theta}^{t+1} \), we check for convergence. The optimization stops if any of the following conditions are met: 1) The change in the optimization variables is sufficiently small, $\Vert\bm{\Theta}^{t+1} - \bm{\Theta}^t\Vert < \epsilon$, where $\epsilon$ is a small positive number. 2) The change in the objective function is sufficiently small: \(|R(\bm{\Theta}^{t+1}) - R(\bm{\Theta}^t)| < \epsilon\).
If the above conditions are met in the $t+1$ iteration, the optimization terminates, and the solution \( \bm{\Theta}^{t+1} \) is considered as the optimal solution.

If the optimization has not converged, we need to update the Hessian matrix for the next iteration. This can be done by the Davidon-Fletcher-Powell (DFP) method, which is used to update the approximate Hessian matrix based on the gradient and step directions, expressed by
\begin{equation}\label{ut1}
	\mathbf{U}^{t + 1} = \mathbf{U}^{t} + \frac{\Delta \bm{\Theta}^{t}[\Delta \bm{\Theta}^{t}]^T}{[\Delta \mathbf{q}^{t}]^T\Delta \bm{\Theta}^{t}} - \frac{\mathbf{U}^{t}\Delta \mathbf{q}^{t}[\Delta \mathbf{q}^{t}]^{T}\mathbf{U}^{t}}{[\Delta \mathbf{q}^{t}]^T\mathbf{U}^{t}\Delta q^{t}},
\end{equation}
where $\Delta \bm{\Theta}^t$ refers to the difference in the decision variables, and $\Delta \mathbf{q}^t$ refers to the difference in the gradient.
Thus the updated Hessian is given by \(\mathbf{U}^{t+1} = \mathbf{U}^t + \Delta \mathbf{U} \), where \( \Delta \mathbf{U} \) is the last two terms in (\ref{ut1}).  

The overall approach is summarized in Algorithm \ref{algo:flexible_precoding}.

\begin{algorithm}[h]
	\caption{Proposed MRA optimization method.}
	\label{algo:flexible_precoding}
	\KwIn{Channel parameters.}
	\KwOut{Optimized antenna positions $\mathbf{x},\mathbf{z},\bm{\psi}^\theta,\bm{\psi}^\phi$.}
	\BlankLine
 
		Define objective function $R(\bm{\Theta})$:
	$
	R(\bm{\Theta}) = -\sum_{k=1}^K \log_2 \left(1 + \frac{P/K}{\sigma^2} \cdot \frac{1}{\left[(\mathbf{H}\mathbf{H}^H)^{-1}\right]_{k,k}} \right)
	$\;
	\For{$t = 1,\cdots,T$}{ 
		Obtain the Taylor expansion based approximation:
	$
		R(\bm{\Theta}^{(t+1)}) \approx \frac{1}{2} \mathcal{S}^{(t),T} \mathbf{U}^{(t)} \mathcal{S}^{(t)} + \mathbf{c}^T \mathcal{S}^{(t)}
		$\;
		Solve QP problem in (\ref{qp})\;
		Update variables: $\bm{\Theta}^{(t+1)} = \bm{\Theta}^{(t)} + \mathcal{S}^{(t)}$\;
		Update $\mathbf{U}^{(t+1)}$ according to (\ref{ut1})\;
		\If{$|\bm{\Theta}^{(t+1)} - \bm{\Theta}^{(t)}| < \epsilon$}{
			\textbf{Break:} Convergence achieved\;
		}
	}  
\end{algorithm}
\section{Simulation Results} 
\begin{figure*} 
	\centering
	\begin{minipage}[b]{0.45\linewidth} 
		\centering
		\includegraphics[width=\linewidth]{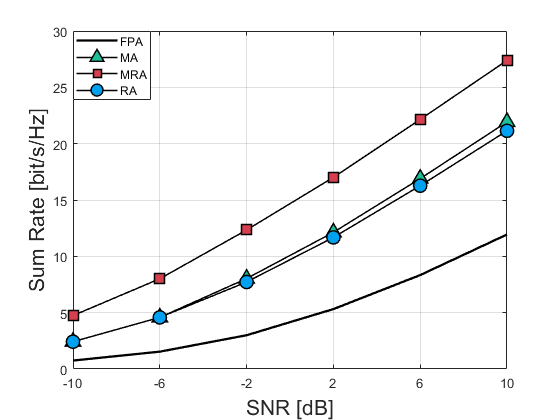}
		\caption{Sum rate varying with SNR ($r=4$,$\psi=\pi/4$).}
		\label{SG}
	\end{minipage}
	\hfill
	\begin{minipage}[b]{0.45\linewidth} 
		\centering
		\includegraphics[width=\linewidth]{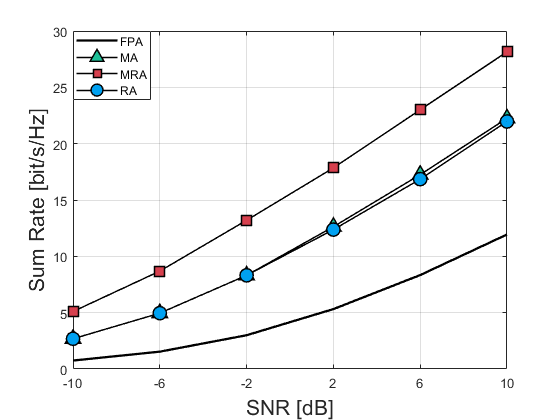}
		\caption{Sum rate varying with SNR ($r=6$,$\psi=\pi/3$).}
		\label{CDF}
	\end{minipage}
	
	\vskip 0.8cm 
	
	\begin{minipage}[b]{0.45\linewidth} 
		\centering
		\includegraphics[width=\linewidth]{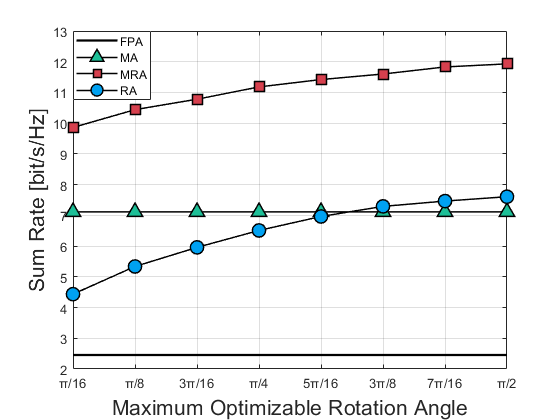}
		\caption{Sum rate varying with $\psi_{\rm max}$.}
		\label{SG2}
	\end{minipage}
	\hfill
	\begin{minipage}[b]{0.45\linewidth} 
		\centering
		\includegraphics[width=\linewidth]{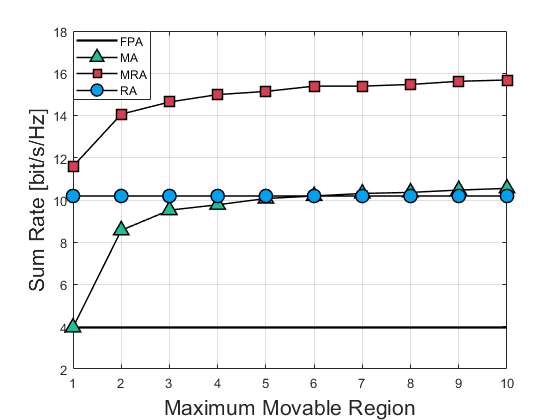}
		\caption{Sum rate varying with $r$.}
		\label{SL2}
	\end{minipage}
	\label{M_L}
\end{figure*}

The system operates at a central frequency of 3 GHz. The BS is equipped with $N = 4$ antennas, serving $K = 4$ users. The user and scatterer locations are uniformly distributed with angles $\theta_{k,l}$ and $\phi_{k,l}$ ranging within [0, 1], and the path gain $\beta_{k,l}$ follows $\mathcal{CN}(0,1)$. All users are assumed to have an identical number of channel paths and equal power. 
It is assumed that all users have the same number of channel paths and transmit power. We define the maximum movable region as $x_{\rm max}=z_{\rm max}= r \frac{\lambda}{2}$, where $r$ is used to control the movable region. We define the maximum rotation angle as $\psi_{\rm max}^\phi=\psi_{\rm max}^\theta =\psi_{\rm max}$.

Four schemes in this paper are considered for evaluation: FPA, MA, Rotatable Antenna (RA), and MRA. The systems performance is evaluated using three key simulation setups: (1) fixing the maximum movable region and the maximum optimizable rotation angle while varying the signal-to-noise ratio (SNR) to compare the performance of the three schemes, (2) fixing the SNR = 1dB and maximum movable region while varying the maximum optimizable rotation angle, and (3) fixing the SNR = 1dB and maximum optimizable rotation angle while varying the maximum movable region.

As can be seen from Fig. \ref{SG}, where the maximum movable range $r = 4$ and the maximum rotation angle $\psi_{\rm max} = \pi / 4$, the performance of all schemes improves with the increase of SNR. FPA performs the worst because it has less degree-of-freedom. The performances of RA and MA are similar in that they both have the ability to change channels to improve the sum rate. Moreover, MRA performs best since it optimizes both antenna position and rotation angle, giving it the most degree-of-freedom. This also shows that optimizing antenna positions and rotation angles can be combined for sum rate enhancement. In Fig. \ref{CDF}, we change the maximum movable range and maximum rotation angle into $r = 6$ and $\psi_{\rm max} = \pi / 3$, respectively. We can observe that similar conclusions can be obtained as shown in Figs. \ref{SG}, demonstrating the potential of antenna position and rotation angle optimization.

Fig. \ref{SG2} presents the sum rate performance when the maximum optimizable rotation angles are varied over \(\{\pi/16, \pi/8, 3\pi/16, \pi/4, 5\pi/16, 3\pi/8, 7\pi/16, \pi/2\}\). The  optimizable rotation angle, \(\psi\), increases, the performance gap between MRA and RA narrows. However, MA and FPA do not show any improvement, as they are not optimized for rotation angles.

In Fig. \ref{SL2}, the sum rate performance of the methods is evaluated with varying maximum movable regions \(r \in \{1, 2, 3, 4, 5, 6, 7, 8, 9, 10\}\). MRA and MA demonstrate a gradual increase in sum rate with increasing \(r\), and the trend becomes smoother as \(G\) increases. In contrast, RA and FPA show no performance change as they are not optimized for the movable region.

\section{Conclusions}\label{Con} 
In this work, we proposed a flexible precoding framework for multi-user MISO systems enhanced by MRAs.  By jointly optimizing antenna positions and rotation angles, we demonstrated significant improvements in sum rate and communication performance compared to traditional fixed-position antenna systems, MAs and RAs. The SQP algorithm is employed to solve the complex, non-linear optimization problem efficiently. Our results highlight the advantages of dynamic antenna configurations in position and orientation, particularly in mitigating interference and enhancing spatial diversity. This approach offers valuable insights for future communication systems, paving the way for broader applications of MRAs in next-generation wireless networks.

\bibliographystyle{IEEEtran}
\bibliography{references.bib}

\begin{thebibliography}{10}
\providecommand{\url}[1]{#1}
\csname url@samestyle\endcsname
\providecommand{\newblock}{\relax}
\providecommand{\bibinfo}[2]{#2}
\providecommand{\BIBentrySTDinterwordspacing}{\spaceskip=0pt\relax}
\providecommand{\BIBentryALTinterwordstretchfactor}{4}
\providecommand{\BIBentryALTinterwordspacing}{\spaceskip=\fontdimen2\font plus
\BIBentryALTinterwordstretchfactor\fontdimen3\font minus
  \fontdimen4\font\relax}
\providecommand{\BIBforeignlanguage}[2]{{%
\expandafter\ifx\csname l@#1\endcsname\relax
\typeout{** WARNING: IEEEtran.bst: No hyphenation pattern has been}%
\typeout{** loaded for the language `#1'. Using the pattern for}%
\typeout{** the default language instead.}%
\else
\language=\csname l@#1\endcsname
\fi
#2}}
\providecommand{\BIBdecl}{\relax}
\BIBdecl

\bibitem{1}
G.~L. Stuber, J.~R. Barry, S.~W. Mclaughlin, Y.~Li, M.~A. Ingram, and T.~G.
  Pratt, ``Broadband mimo-ofdm wireless communications,'' \emph{Proceedings of
  the IEEE}, vol.~92, no.~2, pp. 271--294, 2004.

\bibitem{THZ}
B.~Ning, Z.~Tian, W.~Mei, Z.~Chen, C.~Han, S.~Li, J.~Yuan, and R.~Zhang,
  ``Beamforming technologies for ultra-massive mimo in terahertz
  communications,'' \emph{IEEE Open Journal of the Communications Society},
  vol.~4, pp. 614--658, 2023.

\bibitem{5}
K.~K. Vaigandla and D.~N. Venu, ``Survey on massive mimo: Technology,
  challenges, opportunities and benefits,'' 2021.

\bibitem{6}
L.~Zhu, W.~Ma, and R.~Zhang, ``Modeling and performance analysis for movable
  antenna enabled wireless communications,'' \emph{IEEE Transactions on
  Wireless Communications}, vol.~23, no.~6, pp. 6234--6250, 2024.

\bibitem{FA1}
K.~K. Wong, A.~Shojaeifard, K.-F. Tong, and Y.~Zhang, ``Performance limits of
  fluid antenna systems,'' \emph{IEEE Communications Letters}, vol.~24, no.~11,
  pp. 2469--2472, 2020.

\bibitem{FA2}
K.-K. Wong and K.-F. Tong, ``Fluid antenna multiple access,'' \emph{IEEE
  Transactions on Wireless Communications}, vol.~21, no.~7, pp. 4801--4815,
  2022.

\bibitem{FA3}
H.~{Qin}, W.~{Chen}, Z.~{Li}, Q.~{Wu}, N.~{Cheng}, and F.~{Chen}, ``{Antenna
  Positioning and Beamforming Design for Fluid-Antenna Enabled Multi-user
  Downlink Communications},'' \emph{arXiv e-prints}, p. arXiv:2311.03046, Nov.
  2023.

\bibitem{MA1}
L.~Zhu, W.~Ma, and R.~Zhang, ``Modeling and performance analysis for movable
  antenna enabled wireless communications,'' \emph{IEEE Transactions on
  Wireless Communications}, pp. 1--1, 2023.

\bibitem{MA2}
L.~Zhu, W.~Ma, B.~Ning, and R.~Zhang, ``Movable-antenna enhanced multiuser
  communication via antenna position optimization,'' \emph{IEEE Transactions on
  Wireless Communications}, pp. 1--1, 2023.

\bibitem{MA3}
X.~{Pi}, L.~{Zhu}, Z.~{Xiao}, and R.~{Zhang}, ``{Multiuser Communications with
  Movable-Antenna Base Station Via Antenna Position Optimization},''
  \emph{arXiv e-prints}, p. arXiv:2308.05546, Aug. 2023.

\bibitem{FAA2}
B.~Ning, S.~Yang, Y.~Wu, P.~Wang, W.~Mei, C.~Yuen, and E.~Bj{\"o}rnson,
  ``Movable antenna-enhanced wireless communications: General architectures and
  implementation methods,'' \emph{arXiv preprint arXiv:2407.15448}, 2024.

\bibitem{FA4}
X.~Lai, T.~Wu, J.~Yao, C.~Pan, M.~Elkashlan, and K.-K. Wong, ``On performance
  of fluid antenna system using maximum ratio combining,'' \emph{IEEE
  Communications Letters}, vol.~28, no.~2, pp. 402--406, 2024.

\bibitem{MA5}
S.~Yang, W.~Lyu, B.~Ning, Z.~Zhang, and C.~Yuen, ``Flexible precoding for
  multi-user movable antenna communications,'' \emph{IEEE Wireless
  Communications Letters}, vol.~13, no.~5, pp. 1404--1408, 2024.

\bibitem{MAISAC}
W.~Lyu, S.~Yang, Y.~Xiu, Z.~Zhang, C.~Assi, and C.~Yuen, ``Movable antenna
  enabled integrated sensing and communication,'' \emph{IEEE Transactions on
  Wireless Communications}, pp. 1--1, 2025.

\bibitem{MAISAC2}
Y.~Xiu, S.~Yang, W.~Lyu, P.~L. Yeoh, Y.~Li, and Y.~Ai, ``Movable antenna
  enabled isac beamforming design for low-altitude airborne vehicles,''
  \emph{IEEE Wireless Communications Letters}, pp. 1--1, 2025.

\bibitem{MAMEC}
Y.~{Xiu}, Y.~{Zhao}, S.~{Yang}, M.~{Xu}, D.~{Niyato}, Y.~{Li}, and N.~{Wei},
  ``{Delay Minimization for Movable Antennas-Enabled Anti-Jamming
  Communications With Mobile Edge Computing},'' \emph{arXiv e-prints}, p.
  arXiv:2409.14418, Sep. 2024.

\bibitem{MASECURE}
G.~Hu, Q.~Wu, K.~Xu, J.~Si, and N.~Al-Dhahir, ``Secure wireless communication
  via movable-antenna array,'' \emph{IEEE Signal Processing Letters}, 2024.

\bibitem{12}
Y.~Gao, H.~Vinck, and T.~Kaiser, ``Massive mimo antenna selection: Switching
  architectures, capacity bounds, and optimal antenna selection algorithms,''
  \emph{IEEE Transactions on Signal Processing}, vol.~66, no.~5, pp.
  1346--1360, 2018.

\bibitem{FAA}
S.~{Yang}, J.~{An}, Y.~{Xiu}, W.~{Lyu}, B.~{Ning}, Z.~{Zhang}, M.~{Debbah}, and
  C.~{Yuen}, ``{Flexible Antenna Arrays for Wireless Communications: Modeling
  and Performance Evaluation},'' \emph{arXiv e-prints}, p. arXiv:2407.04944,
  Jul. 2024.

\end{thebibliography}

\vspace{12pt}

\end{document}